\documentclass[twocolumn]{aastex631} 

\shorttitle{}
\shortauthors{Ma et al.}
\graphicspath{{./}{figures/}}

\usepackage{amsmath}
\usepackage{threeparttable}

\begin{document}

\title{Digitization of Astronomical Photographic Plate of China: Photometric Calibration of Single-exposure Plates and Release of Photometric Catalogs}
\correspondingauthor{Haibo Yuan, Kai Xiao}
\email{yuanhb@bnu.edu.cn, xiaokai@ucas.ac.cn}

\author[0009-0003-1069-1248]{Mingyang Ma}
\affiliation{Institute for Frontiers in Astronomy and Astrophysics, Beijing Normal University, Beijing, 102206, China}
\affiliation{School of Physics and Astronomy, Beijing Normal University, Beijing, 100875, China}

\author[0000-0003-2471-2363]{Haibo Yuan}
\affiliation{Institute for Frontiers in Astronomy and Astrophysics, Beijing Normal University, Beijing, 102206, China}
\affiliation{School of Physics and Astronomy, Beijing Normal University, Beijing, 100875, China}

\author[0000-0001-8424-1079]{Kai Xiao}
\affiliation{School of Astronomy and Space Science, University of Chinese Academy of Sciences, Beijing 100049, China}
\affiliation{Institute for Frontiers in Astronomy and Astrophysics, Beijing Normal University, Beijing, 102206, China}

\author[0000-0002-1259-0517]{Bowen Huang}
\affiliation{Institute for Frontiers in Astronomy and Astrophysics, Beijing Normal University, Beijing, 102206, China}
\affiliation{School of Physics and Astronomy, Beijing Normal University, Beijing, 100875, China}

\author[0000-0002-4878-1227]{Tao Wang}
\affiliation{Institute for Frontiers in Astronomy and Astrophysics, Beijing Normal University, Beijing, 102206, China}
\affiliation{School of Physics and Astronomy, Beijing Normal University, Beijing, 100875, China}

\author[0000-0003-3250-2876]{Yang Huang}
\affiliation{School of Astronomy and Space Science, University of Chinese Academy of Sciences, Beijing 100049, China}
\affiliation{CAS Key Lab of Optical Astronomy, National Astronomical Observatories, Chinese Academy of Sciences, Beijing 100012, China}

\author{Shiyin Shen}{}
\affiliation{Shanghai Astronomical Observatory, Chinese Academy of Sciences, Shanghai 200030, China}
\author{Zhengjun Shang}{}
\affiliation{Shanghai Astronomical Observatory, Chinese Academy of Sciences, Shanghai 200030, China}
\author{Yong Yu}{}
\affiliation{Shanghai Astronomical Observatory, Chinese Academy of Sciences, Shanghai 200030, China}
\affiliation{School of Astronomy and Space Science, University of Chinese Academy of Sciences, Beijing 100049, China}
\author{Liangliang Wang}{}
\affiliation{Shanghai Astronomical Observatory, Chinese Academy of Sciences, Shanghai 200030, China}
\author{Meiting Yang}{}
\affiliation{Shanghai Astronomical Observatory, Chinese Academy of Sciences, Shanghai 200030, China}
\author{Jing Yang}{}
\affiliation{Shanghai Astronomical Observatory, Chinese Academy of Sciences, Shanghai 200030, China}
\author{Min Liu}{}
\affiliation{Shanghai Astronomical Observatory, Chinese Academy of Sciences, Shanghai 200030, China}
\author{Quanfeng Xu}{}
\affiliation{Shanghai Astronomical Observatory, Chinese Academy of Sciences, Shanghai 200030, China}
\affiliation{School of Astronomy and Space Science, University of Chinese Academy of Sciences, Beijing 100049, China}
\author{Chenzhou Cui}{}
\affiliation{CAS Key Lab of Optical Astronomy, National Astronomical Observatories, Chinese Academy of Sciences, Beijing 100012, China}
\affiliation{National Astronomical Data Center, Chinese Academy of Sciences, Beijing 100101, China}
\author{Dongwei Fan}{}
\affiliation{CAS Key Lab of Optical Astronomy, National Astronomical Observatories, Chinese Academy of Sciences, Beijing 100012, China}
\affiliation{National Astronomical Data Center, Chinese Academy of Sciences, Beijing 100101, China}
\author{Zhenghong Tang}{}
\affiliation{Shanghai Astronomical Observatory, Chinese Academy of Sciences, Shanghai 200030, China}
\affiliation{School of Astronomy and Space Science, University of Chinese Academy of Sciences, Beijing 100049, China}
\author{Jianhai Zhao}{}
\affiliation{Shanghai Astronomical Observatory, Chinese Academy of Sciences, Shanghai 200030, China}

\begin{abstract}
The Chinese Plate-Digitizing Project has digitized a total number of about 30,000 
astronomical plates from 11 telescopes of five observatories (SHAO, NAOC, PMO, YNAO, and QDO) in China, spanning nearly 100 years of observations. 
In this work, we present a photometric calibration method to calibrate about 15,000 single-exposure plates to the JKC photometric system.
Using standard stars constructed from the BEST database, we have identified and corrected various systematic effects (including the magnitude term, color term, and flat-field term) to a high precision.
The final photometric precision is typically 0.15, 0.23, 0.17, 0.11, 0.19 mag for plates collected in SHAO, NAOC, PMO, YNAO, and QDO, respectively, with best cases reaching 0.07, 0.08, 0.06, 0.05, and 0.11 mag, respectively. Candidates of variable sources are also identified. The catalogs are publicly available at the China National Astronomical Data Center.  
Such a dataset will provide a valuable resource for conducting long-term, temporal-scale astronomical research.
Our calibration method can also be applied to other digitized astronomical plates in general.
\end{abstract}


\section{Introduction} \label{sec:intro}

Photographic plates are the primary means of recording celestial images in the 19th and 20th centuries, before widespread use of charge-coupled devices (CCDS) and other photonic electron detectors. Currently, there are more than 10 million photographic plates preserved worldwide (\citealt{2018AN....339..408H}; \citealt{2019AN....340..690H}).
These images have contributed to a number of important discoveries (e.g., \citealt{1985AJ.....90.2474A}; \citealt{1989ApJ...336..889H}; \citealt{1990A&A...235..174H}; \citealt{2000AJ....119.1914T}). Even today, with more and more accurate observations, the information in these photos is still important (e.g., \citealt{2021MNRAS.508.3649D}; \citealt{2022MNRAS.516.2095Y}; \citealt{2023A&A...680A..41P}), as they are the only way to study the behavior of objects over very long time intervals (100 years or more). 

With the passing of time, the quality of the photographic plates has gradually deteriorated. To permanently preserve these precious historical heritages, digitization has proven to be an effective method. 
Countries around the world have made efforts to digitize the astronomical plates in their collections, such as the SSS project in Great Britain \citep{2001MNRAS.326.1279H}, the D4A project in Belgium \citep{2003ASPC..295...93D}, the DASCH project in the United States \citep{2009ASPC..410..101G}, the APPLAUSE project in Germany \citep{2024A&A...687A.165E}, and the NAROO project in France \citep{2021A&A...652A...3R}.

Chinese Plate-Digitizing Project began in 2009 (\citealt{2007PABei..25....1J}). About 30,000 astronomical photographic plates were digitized between 2012 and 2017 with a special digitizing machine that has high precision in both astrometry and photometry (\citealt{2016MNRAS.457.2900Y}) by the Shanghai Plate Digitization Lab (\citealt{2017RAA....17...28Y}; \citealt{2019IAUS..339...69T}). Recently, astrometric calibration was carried out on 15,696 conventional single-exposure imaging plates\footnote{\raisebox{-6.6ex} {\parbox{7.9cm}{Note that there are also several thousand multiple-exposure imaging plates digitized by the Chinese Plate-Digitizing Project. 
Multiple-exposure plates refer to those when multiple exposures of the observed sky were recorded on the same plate by slightly adjusting the telescope’s pointing for economic reasons.}}}
(\citealt{2024RAA....24e5010S}). The plates come from eleven telescopes at five observing sites, covering an observation period of 99 years (1901--1999).
Images of various types of objects were recorded in these plates, including asteroids, comets, binary stars, variable stars, eruptive stars, radio stars, star clusters, nebulae, and extragalactic objects.
All the plate images, together with plate information and measured coordinates of all the objects on the plates, have been archived at the National Astronomical Data Center (NADC)\footnote{\url{https://nadc.china-vo.org/res/r100742/}}. More detailed information of these plates and their digitization processes can be found in \cite{2024RAA....24e5010S} and references therein.

The photometric calibration of digitized photo-plates is essential but challenging. Compared with the widely used CCD detectors nowadays, the difficulties in calibrating  photographic plates include but are not limited to: 1) very strong non-linearity in the response of the plates; The response curve of photographic plates is
traditionally described by a characteristic curve of log (Exposure) versus log (Density), which is typically S-shaped, with an approximate linear portion in the middle. 2) very large field of view and complex flat-field structure; 3) varying quality of the plates; 4) susceptibility of the plates to many random factors during the development process, such as the formulation, time, and temperature of the development; Therefore, each plate has to be calibrated individually and independently. 5) plates with emulsion delamination, mold, damage, etc.

Thanks to the Gaia data release 3 (DR3; \citealt{2023A&A...674A...1G}), 
flux-calibrated low-resolution BP/RP (XP) spectra for 220 million sources in the wavelength range $330{\rm \,nm}\leq\lambda\leq 1050{\rm \,nm}$ are available, with the majority having $G<17.65$. 
The Gaia DR3 XP spectra allow us to synthesize the magnitudes of stars by mapping the spectral energy distribution of stars at the top of the Earth's atmosphere to any passbands covered by the Gaia XP spectra in the specific photometric system. Although the Gaia DR3 XP spectra contain complex systematic errors  (e.g., magnitude- and color-dependent; \citealt{2023A&A...674A...3M}), particularly at wavelengths below 400 nm, \cite{2024ApJS..271...13H} have carefully measured and precisely corrected these systematics. 
Following \cite{2012PASP..124..140B} and \cite{2023A&A...674A..33G}, \cite{2023ApJS..269...58X} developed the synthetic photometry method based on the ``corrected" Gaia DR3 XP spectra \citep{2024ApJS..271...13H} , which has achieved notable success in high-precision photometric calibration for wide-field photometric surveys.  For example, \cite{2023ApJS..269...58X}, \cite{2024ApJS..271...41X} and \cite{2025ApJS..277...26L} have re-calibrated the J-PLUS DR3, S-PLUS DR4, and USS DR1 photometric data to a precision of 1 to 6 mmag in their zero-points.
The synthetic photometry method based on ``corrected" Gaia DR3 XP spectra naturally enables the construction of approximately 200 million high-precision photometric standard stars, providing high-quality calibration reference stars for the precise photometric calibration of digitized photographic plates.

The manuscript is formatted as follows. We present the data used in this work in Section \ref{sec:data}, followed by a description of the calibration method and process in Section \ref{sec:method}. We present and discuss the calibration results in Section \ref{sec:results}. The calibrated photometric catalogs and candidates of variable sources are introduced in Section \ref{sec:catalog}. Finally, we conclude in Section \ref{sec:conclusion}.

\section{Data} \label{sec:data}
\subsection{Digitized Chinese Photographic Plates} \label{sec:photo}

A total of $15,696$ Chinese single-exposure plates were digitized and underwent astrometric calibration.
Among them, there are $1,539$ plates from Shanghai Astronomical Observatory (SHAO), $4,980$ from the National Astronomical Observatory of the Chinese Academy of Sciences (NAOC), $8,283$ from Purple Mountain Observatory (PMO), $753$ from Yunnan Astronomical Observatory (YNAO), and $141$ from Qingdao Observatory (QDO).
Astrometric processing of plates includes star extraction using SExtractor \citep{1996A&AS..117..393B}, parameter model of plates and the celestial coordinates of all stars computation using Astrometry.net \citep{2010AJ....139.1782L}, and celestial coordinate calculation using reference star of Gaia data release 2 (DR2; \citealt{2018A&A...616A...1G}). Finally, for long-focus telescopes, the astrometric accuracy of the plates ranges from 0.1{\arcsec} to 0.3{\arcsec}, while for medium- and short-focus telescopes, it ranges from 0.3{\arcsec} to 1.0{\arcsec}. This resulted in a catalog containing 43,321,336 unique common sources with Gaia DR2, comprising 290,417,004 individual observations, with some sources being observed multiple times \citep{2024RAA....24e5010S}.

This catalog provides object name (\texttt{OBJNAM}), field of view, observational date, photographic plate type (\texttt{EMULS}), and filter type (\texttt{FILT}) for some, rather than all, plates. For example, 18\% of the plates lack records of their plate types, and 54\% lack records of filters. Additionally, the catalog includes, but is not limited to, the Right Ascension (\texttt{RA}) and Declination (\texttt{DEC}) of the sources, their positions (\texttt{XWIN\_IMAGE} and \texttt{YWIN\_IMAGE}) on the photographic plates, instrumental magnitudes (\texttt{MAG\_AUTO}), and \texttt{FLAGS} used to assess the photometric quality.
Note that only common sources with Gaia DR2 were used in our work, as other targets were heavily contaminated.

\subsection{Photometric Standard Stars} \label{sec:stdstar}

As part of the BEst STar (BEST) database (K. Xiao et al., in preparation), over 200 million high-precision all-sky photometric standard stars, brighter than approximately $G$=17.65 mag,  were constructed for nearly 20 photometric systems (e.g., JKC, PS1, and SDSS) using the ``corrected" Gaia DR3 XP spectra \citep{2024ApJS..271...13H}.

In this study, we used the $UBVRI$ standard photometry from the BEST catalog within the classic JKC system \citep{2012PASP..124..140B} as the photometric standard stars, given that the JKC system has been widely used for decades. Note that the JKC standard stars in the BEST database have been standardized to the Landolt standard stars \citep{2013AJ....146...88C}. This approach offers a reasonable approximation in the absence of detailed photometric system information. Additionally, we filtered the standard stars by excluding those where the third digit of the \texttt{Syn\_FLAG}, which indicates the reliability of the Gaia DR3 XP spectra correction \citep{2024ApJS..271...13H}, was less than 3. 
The standard stars with \texttt{excess\_factor} greater than $1.3+0.06 (BP-RP)^2$ \citep{2022ApJS..258...44X} were also filtered out.

\section{Method} \label{sec:method}

For each plate, the calibration process consists of four steps: 1) constructing standard star samples; 2) selecting the closest JKC passband; 3) removing false stars; and 4) calculating zero-points. In the following subsections, we will describe these steps, respectively.

\subsection{Constructing standard star samples} \label{sec:process1}
We matched the photometric standard stars (Section \ref{sec:stdstar}) to construct a calibration sample having JKC $UBVRI$ standard magnitudes, using a cross-matching radius of $1^{\prime\prime}$

\subsection{Selecting the closest JKC passband} \label{sec:process2}
There is a wide variety of photographic plates with different sensitive wavelength ranges, see Table\,\ref{tab:kodak}. For the convenient use of the calibrated catalogs, we chose to calibrate each plate to its closest JKC photometric system.
However, due to the fact that the plate/filter types are missing for a significant fraction of plates (about 18\%/54\%), we chose to select the closest band empirically for all the plates. The details are as follows.

\begin{figure*}[ht] \centering
\includegraphics[width=14.5cm]{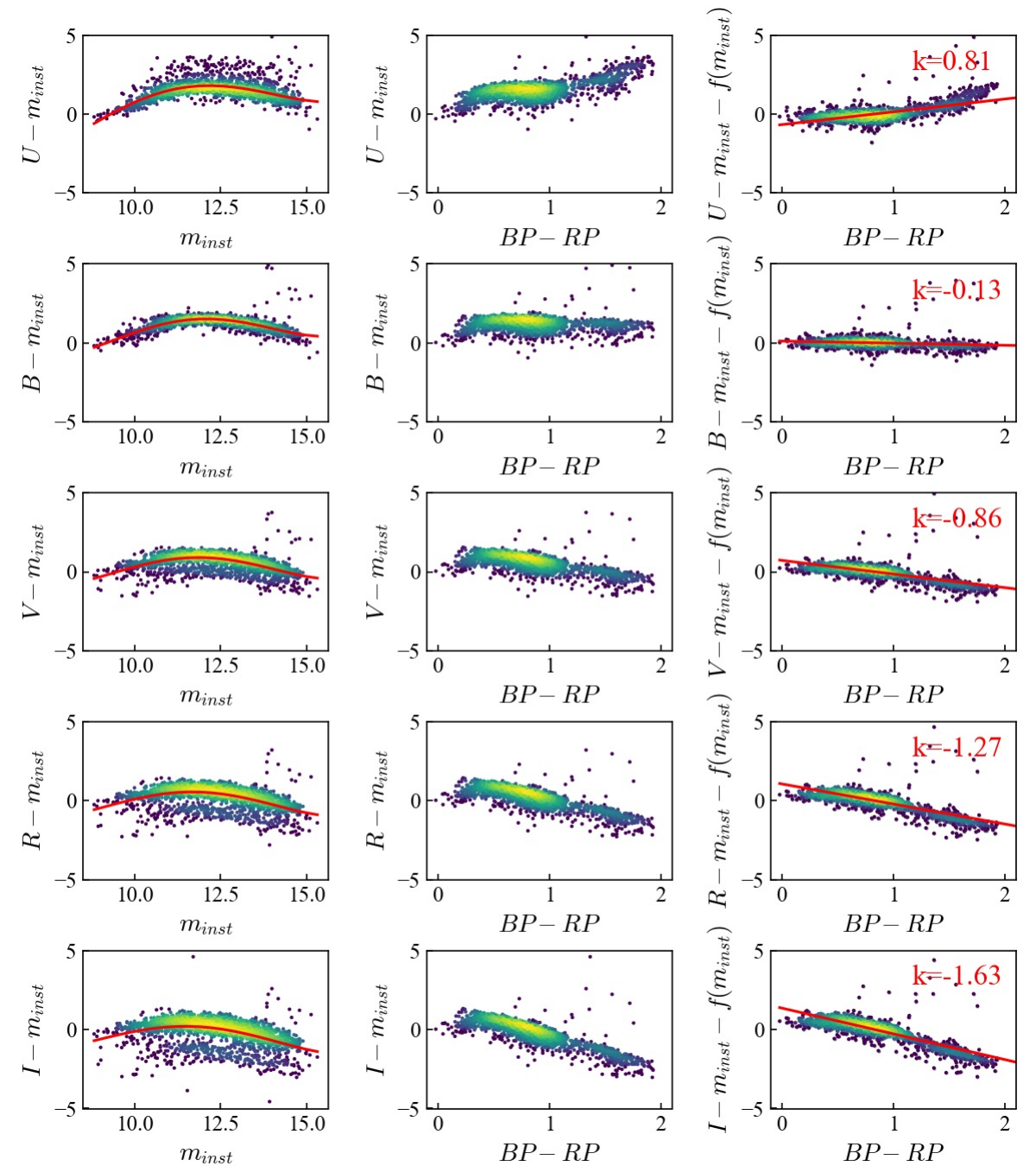} 
\caption{An example (SH1301382001 from SHAO) to demonstrate the selection process of the closest JKC passband. The rows from top to bottom are for results of the $U/B/V/R/I$ bands, respectively.  For each row, the left panel shows the magnitude differences as a function of instrumental magnitude, where the red line indicates the polynomial fitting result. The middle panel shows the magnitude differences as a function of \(BP-RP\). The right panel shows the fitting residuals as a function of \(BP-RP\), where k represents the slope of the linear regression.}
\label{Fig:wave}
\end{figure*}

For each plate, we first corrected for its magnitude-related systematic errors for each JKC band using a second-order polynomial. Then, we calculated the slopes of the fitting residuals versus \(BP-RP\). The band with the flattest slope was selected as the closest band. 
One example is shown in Figure\,\ref{Fig:wave}. The \(B\) band with slope of $-0.13$ was selected as the closest passband. For a plate with the type \(O\), this is
as expected (see Table\,\ref{tab:kodak}).

The success rate for finding the closest band is $85\%$ for plates from YNAO and $92-98\%$ for plates from other observatories. The success rate to find the first two closest bands increases to $94\%$ for plates from YNAO and $99-100\%$ for plates from other observatories. 

\begin{deluxetable*}{lcccccccccc}[ht!]
\tablecaption{Sensitive wavelength ranges of different Kodak plates and closest JKC bands\label{tab:kodak}}
\tablehead{
Kodak plate type & \colhead{${O}$} & \colhead{${J}$} & \colhead{${G}$} & \colhead{${D}$} & \colhead{${E}$} & \colhead{${F}$} & \colhead{${N}$}}
\startdata
Wavelength\ range (Å)& $3000\sim5000$ & $4500\sim5500$ & $4500\sim5800$ & $4500\sim6300$ & $5600\sim6900$  & $4500\sim6900$ & $6800\sim8900$ \\
JKC\ bands$^a$ & ${B/U}$& ${B/V}$ & ${B/V}$ & ${V/B}$ & ${R/V}$ & ${V/R}$ & ${R/I}$
\enddata
\begin{tablenotes}[para, flushleft]
\item $^a$ For a given plate type, two closest JKC bands are given in order.
\end{tablenotes}
\end{deluxetable*}

\subsection{Removing false stars} \label{sec:process3}

\begin{figure*}[ht!] \centering
\includegraphics[width=16 cm]{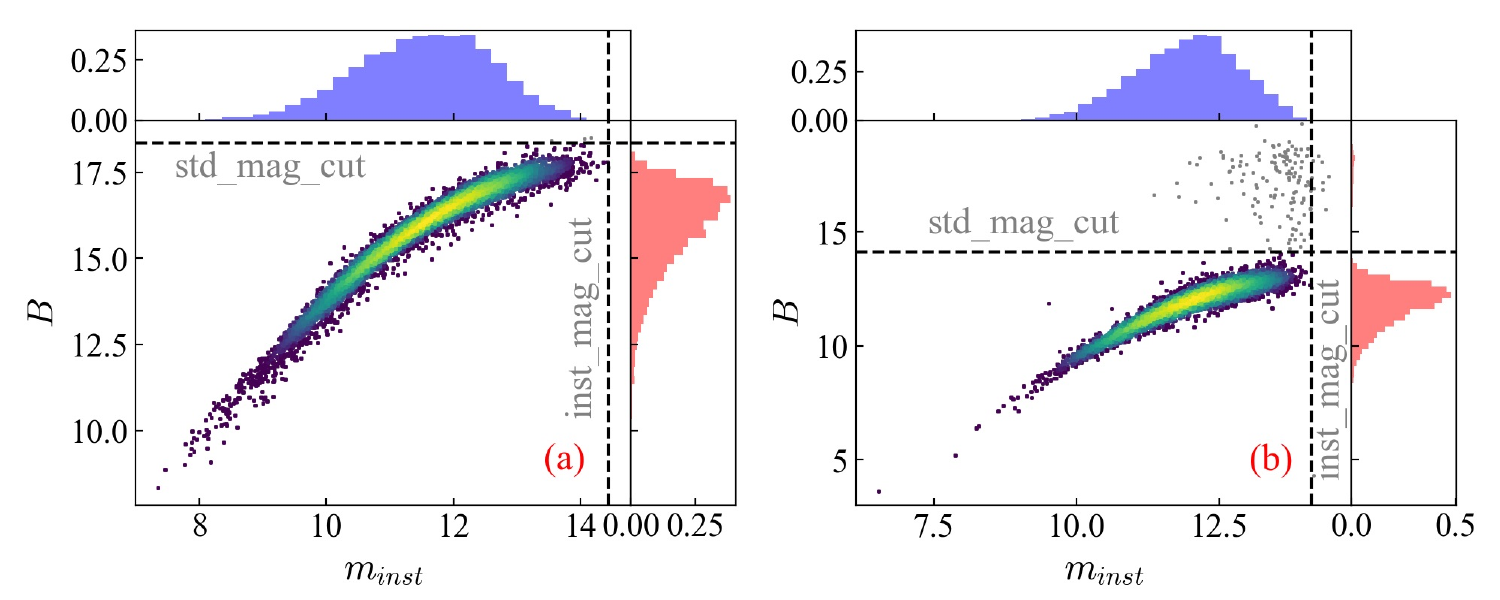} 
\caption{Examples of removing false stars. Panels (a--b) are for plates SH6001G60119001 and BJ7004DA757k001, respectively. The horizontal and vertical dashed lines indicate the cuts on $std\_mag\_cut$ and $inst\_mag\_cut$, respectively.}
\label{Fig:cut}
\end{figure*}

In the standard magnitude--instrumental magnitude diagram, most stars on one plate are distributed along a smooth curved line, as shown in Figure\,\ref{Fig:cut} (a).
However, for certain plates, many stars are located above the curve. These stars are false positives generated by noise on the plates because they primarily originate from plates with deep fields or dense fields. Therefore, these false sources are unsuitable for calibration.
 
The $std\_mag\_cut$ is set to exclude false positive stars, as shown in Figure\,\ref{Fig:cut} (b). And $inst\_mag\_cut$ are used to exclude stars with excessively large instrumental magnitudes, as these stars are located far from the curve and have large measurement errors.  

For each plate, we bin the standard magnitudes at regular intervals, as shown in the red histogram, and count the number of stars in each bin from the bright end to the faint end. When the number of stars in a bin first drops below a certain percentage, the median standard magnitude of that bin is taken as the upper threshold ($std\_mag\_cut$) for the standard magnitude. The method for obtaining $inst\_mag\_cut$ is similar to that of $std\_mag\_cut$, after  applying the cut on $std\_mag\_cut$.

The bin width and the percentage mentioned above were determined empirically and may vary between different plates. Therefore, after calibrating all plates, we manually selected and re-calibrated those plates (approximately $400$) where the choices for $std\_mag\_cut$ and $inst\_mag\_cut$ were clearly unreasonable. Note that to ensure that calibration proceeded smoothly, plates with fewer than $10$ standard stars were excluded in the calibration. And for plates with fewer than $30$ standard stars, no stars were excluded in this step. 

\subsection{Calculating zero-points} \label{sec:process4}

Photometric calibration in this work is to find the relationship between the instrumental magnitude and the selected standard magnitude. We consider three effects: 1) non-linear response of the photographic plates, expressed as magnitude term; 2) difference between the real photometric system and the adopted JKC system, expressed as color term; and 3) flat field to correct any position-dependent systematic errors. 
We determined the zero-point for each plate individually, modeling it independently as functions of the instrumental magnitude ($m_{\rm inst}$), Gaia $BP-RP$ color\footnote{\raisebox{-6.6ex} {\parbox{7.9cm}{There are two reasons for using today's \(BP-RP\) here: 1) The change in color is usually much smaller than the change in brightness for a given target; 2) The color term is relatively flat, indicating that even if there is a deviation between today's \(BP-RP\) and historical \(BP-RP\), the impact is not significant.}}}, and the source's position in the plate ($X$ and $Y$). Therefore, the calibrated magnitude ($m$) can be expressed as:

\begin{equation}
\begin{aligned}
{m_{\rm inst}} + f_m(m_{\rm inst}) + f_c(BP-RP) + f_p(X,~Y)~.
\end{aligned}
\end{equation}

Here, the magnitude and color terms, denoted as $f_m$ and $f_c$, are expressed as a fourth- and second-order polynomial, respectively, and are defined as follows:

\begin{equation}
\begin{aligned}
f_m = \sum_{i=0}^{4} a_{i} m_{\rm inst}^{i},~~f_c = \sum_{i=0}^{2} b_{i} (BP-RP)^{i}~.
\end{aligned}
\end{equation}
Where, $a$ and $b$ are sets of constants.

The position-dependent term, $f_p$, consists of two parts: $f_{ls}$ and $f_{ss}$. The former is a third-order two-dimensional polynomial (with 10 free parameters; refer to Equation\,\ref{equation:3}) as a function of $X$ and $Y$. It is used to fit the large-scale spatial structure. The latter 
is used to describe the complex and relatively small-scale spatial structures. The similar structures are widely presented in modern wide-field photometric surveys; more details can be found in \citep{2023ApJS..269...58X,2024ApJS..271...41X,2023arXiv230713238X,2023ApJS..268...53X}.

\begin{equation}
\begin{aligned}
f_{ls}(X,~Y)=\sum_{j=0}^{3} \sum_{i=0}^{j}p_{i, j-i} X^{i} Y^{j-i}~. \label{equation:3}
\end{aligned}
\end{equation}

It is worth noting that the number of standard stars is unevenly distributed with $m_{\rm inst}$ and $BP-RP$.
In order to achieve a robust polynomial fitting of $f_m$ and $f_c$, we assigned higher weights to stars within less populated regions, with the weight inversely proportional to the square root of the stellar density.

To determine $f_{ss}$, we adopted the improved 
numerical stellar flat-field correction method of \cite{2024ApJS..271...41X}. 
The idea behind this method is to predict the $f_{ss}$ of the source ($X_{0},~Y_{0}$) according to its neighboring standard stars using a low-order polynomial as the kernel function. In this work, we adopted a first-order two-dimensional polynomial, as described by Equation\,\ref{equation:4},

\begin{equation}
 \begin{aligned}
  f_{ss}(X,~Y) = q_{0} X + q_{1} Y + q_{2}~.\label{equation:4}
 \end{aligned}
\end{equation}
where, $q_{0}$, $q_{1}$ and $q_{2}$ are constants related to the source's pixel coordinate. 

It is crucial to select a reasonable neighborhood radius (in units of pixel) that balances its small size with an sufficient number of standard stars.
For each plate, we first set a threshold value of the number of neighboring standard stars ($N_{\rm std}^{\rm TS}$, see Equation\,\ref{equation:5}) according to the number of standard stars on the plate ($N_{\rm std}$). Then, an initial value for the neighborhood radius, $a$, is estimated as $\sqrt{(X_{\rm max}-X_{\rm min})(Y_{\rm max}-Y_{\rm min}) N_{\rm std}^{\rm TS} / N_{\rm std}}$. 
For each star in a given plate, if the number of neighboring standard stars with a radius of $a$ is less than $N_{\rm std}^{\rm TS}$, the neighborhood radius will increase by $a$ until the number of neighboring standard stars is no less than $N_{\rm std}^{\rm TS}$ or the neighborhood radius reaches $10a$. 

\begin{equation} \label{equation:5}
\begin{aligned}
N_{\rm std}^{\rm TS} = 
\begin{cases}
    N_{\rm std} / 3, & \text{if } N_{\rm std} \leq 45 \\
    15, & \text{if } 45 < N_{\rm std} \leq 3375 \\
    N_{\rm std}^{1/3}, & \text{if } 3375 < N_{\rm std} \leq 64000 \\
    40, & \text{if } 64000 < N_{\rm std}
\end{cases}
\end{aligned}
\end{equation}

\begin{figure}[ht!] \centering
\includegraphics[width=7.8 cm]{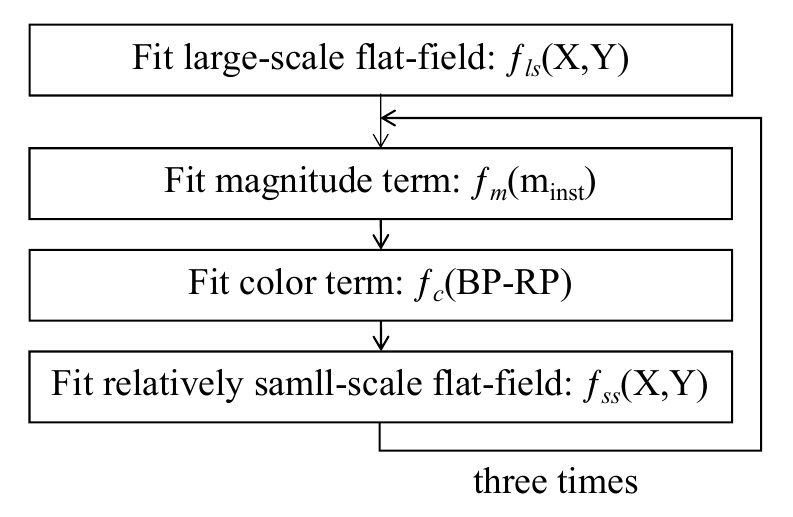} 
\caption{Flowchart to estimate zero-point of each plate.}
\label{Fig:calibrate}
\end{figure}

We adopted an iterative approach to estimate different terms of the zero-points, as shown in Figure \ref{Fig:calibrate}. The large-scale flat-field structure was corrected prior to the iteration process. Good and convergent calibration results can be obtained after three iterations. 

\begin{figure*}[ht!] \centering
\includegraphics[width=17.5 cm]{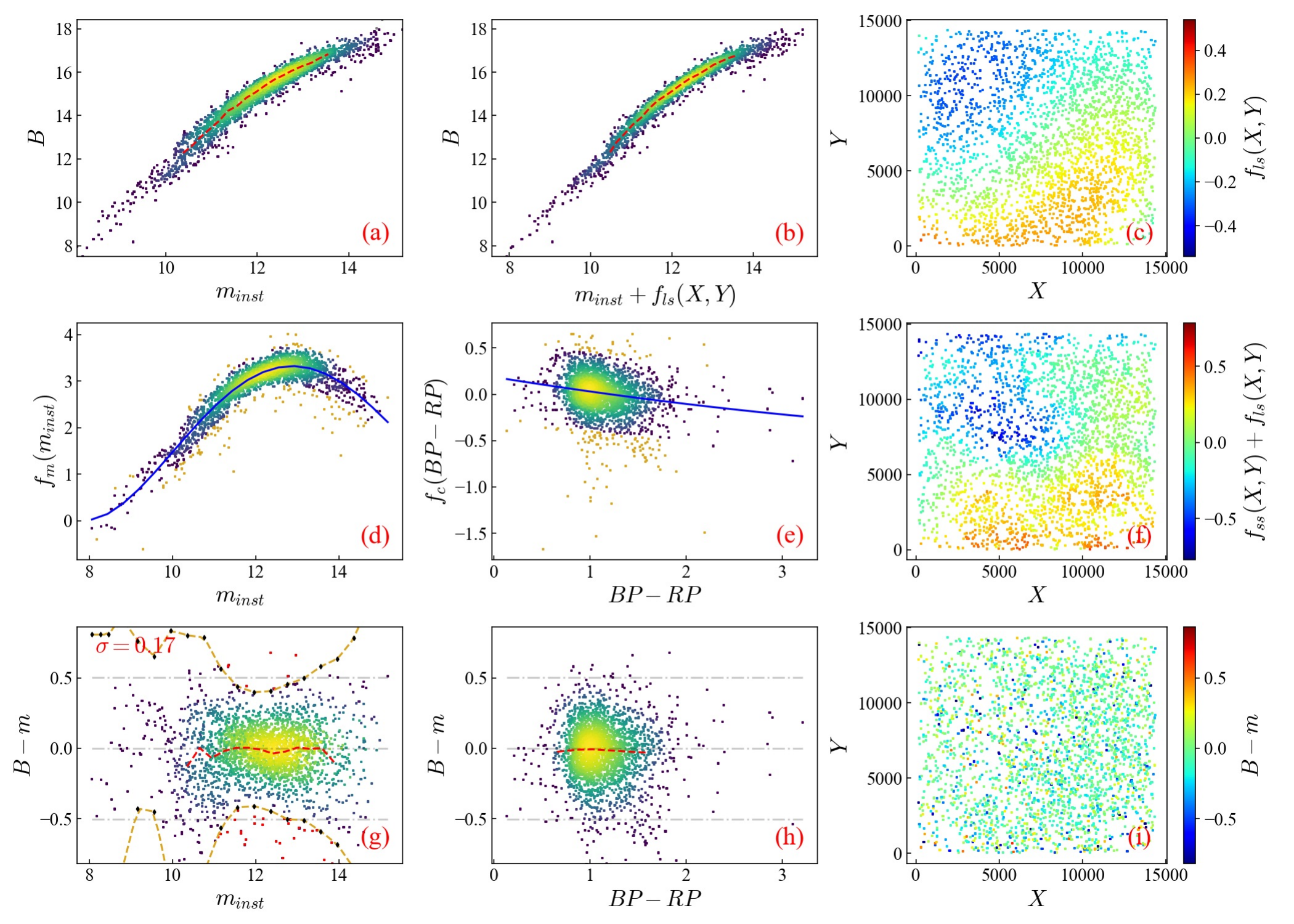}
\caption{Calibration results of QD830883-20001. Panels (a--b) shows instrumental magnitude versus standard magnitude before and after large-scale flat field correction, respectively. The distribution is much tighter after the correction. Panel (c) displays the large-scale flat field. Panels (d--f) plot the final magnitude-, color-, and flat-field terms, respectively. The blue curves in panels (d--e) are the fitted polynomials. 
The relatively small-scale structure is visible in panel (f).
The overall flat-field structure closely resembles that of Figure\,\ref{Fig:flat}.
Panels (g--i) show the differences between the calibrated magnitudes and the standard magnitudes as a function of $m_{inst}$, \(BP-RP\), and plate coordinates, respectively. The red dashed lines in panels (g--h) indicate the median differences, no further trends are found. The yellow dashed lines in panel (g) indicate the $3\sigma$ standard deviations. The overall standard deviation is also marked. The red dots are candidates of variable stars. }
\label{Fig:example}
\end{figure*}

\begin{figure}[ht!] \centering
\includegraphics[width=7.5 cm]{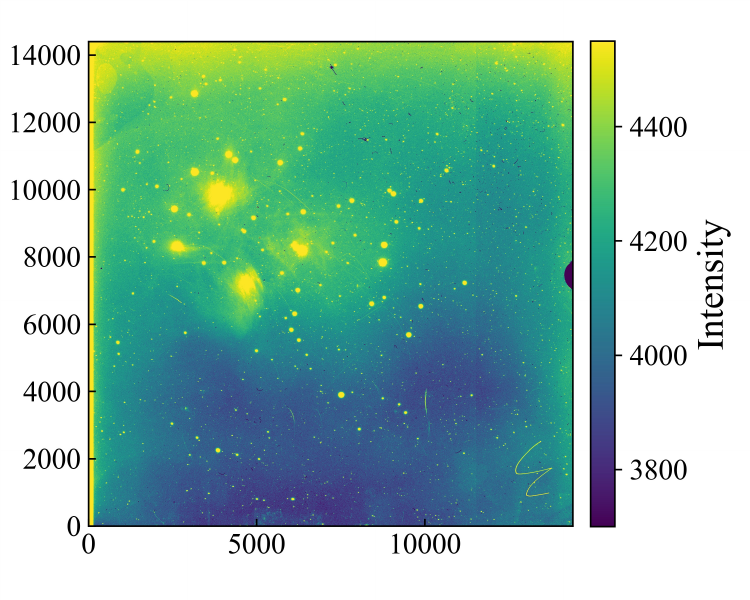} 
\caption{Digitized image of QD830883-20001.}
\label{Fig:flat}
\end{figure}

\section{Results}\label{sec:results}

In this section, we present and discuss the calibration results. A total number of 15,680 (99.9\%) plates have been calibrated.
The remaining 16 plates have too few (less than 10) standard stars to calibrate. 
An example of calibration results from QDO is shown in Figure \ref{Fig:example}. More examples from other four stations are presented in Figures\,\ref{Fig:sh}, \ref{Fig:bj}, \ref{Fig:zt}, \ref{Fig:yn} in the appendix. The figures show that we have successfully corrected the magnitude-, color-, and flat-field terms to a high precision.

\begin{figure*}[ht!] \centering
\resizebox{\hsize}{!}{\includegraphics{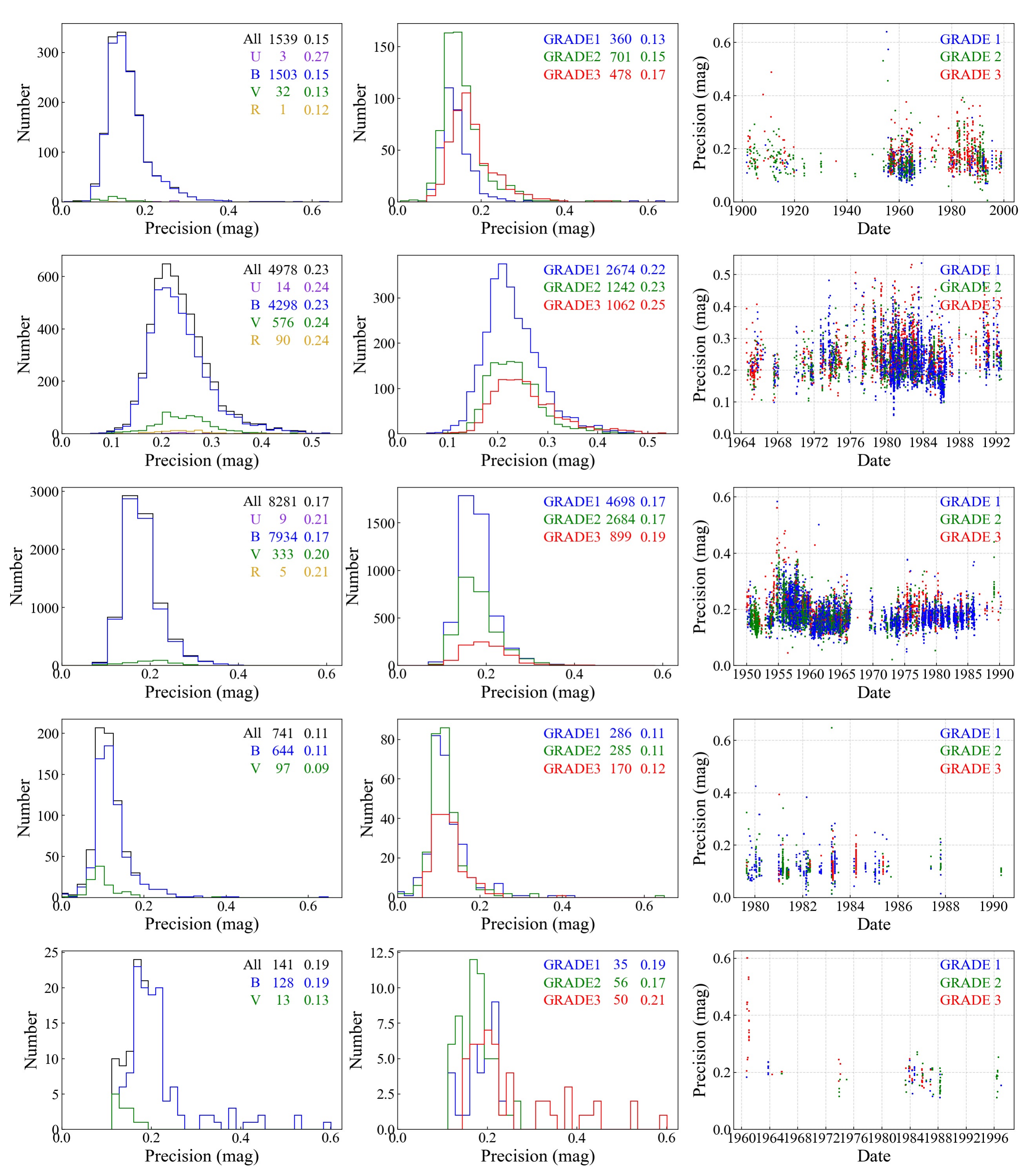}} 
\caption{The calibration statistics. From top to bottom are for the SHAO, NAOC, PMO, YNO, and QDO plates, respectively. The left and middle panels show the histogram distributions of photometric precision of different filters and plate grades, respectively. The number of plates and median precision are labeled. The right panels plot precision as a function of observational time.}
\label{Fig:Statistics}
\end{figure*}

By comparing with the Gaia standard stars, the final photometric precision was estimated for each plate. The typical values are 0.15, 0.23, 0.17, 0.11, 0.19 mag for the SHAO, NAOC, PMO, YNAO, and QDO plates, respectively, with best cases reaching 0.07, 0.08, 0.06, 0.05, and 0.11 mag, respectively.
The left panels of Figure\,\ref{Fig:Statistics} show the histogram distributions of photometric precision of different filters. No strong dependence on filters is found. The middle and right panels show that plates of larger grade (lower quality) usually suffer lower precision.

The limiting magnitude of each plate was estimated as the magnitude of the faintest $5\%$ of all calibrated stars. The typical limiting magnitude ranges are 13.4 -- 17.7, 13.7 -- 18.5, 13.6 -- 18.0, 12.2 -- 17.5, 12.7 -- 18.3 mag for the SHAO, NAOC, YNAO, QDO, and PMO plates, respectively. Most photographic plates in this work are calibrated to the B or V bands. The 3-sigma limiting magnitudes of the B and V bands in the BEST catalog are 19.0 and 19.7 mag, respectively, which are sufficient to cover all the stars in the Chinese digital astronomical plates.

\begin{figure*}[ht!] \centering
\includegraphics[width=11.cm]{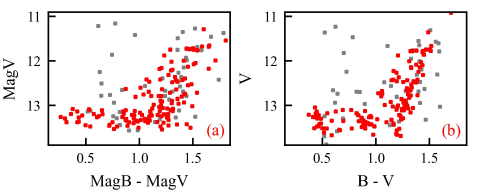} 
\caption{Color-magnitude diagram of globular cluster M\,4. The $MagB$ and $MagV$ magnitudes are from the calibrated catalogs; the $B$ and $V$ magnitudes are from Gaia DR3. The red dots represent the member stars, and the gray dots are the other field stars.}
\label{Fig:m4}
\end{figure*}

To further validate our calibration results, Figure\,\ref{Fig:m4} (a) shows the color--magnitude diagram of the globular cluster M\,4 using data from two YNAO plates: YN8105372001 and YN8305682001, with photometric precision of $0.08$ and $0.07 {\rm \,mag}$, respectively. For comparison, the Gaia DR3 result is plotted in Figure\,\ref{Fig:m4} (b). 
The red dots represent member stars of M\,4, and the gray dots are the other field stars. 
The features are similar in both panels, suggesting that our photometry and calibration is reliable.  
Additionally, we cross-matched the Landolt standard stars \citep{2013AJ....146...88C} with our catalogs, yielding about 4000 common stars. Most common stars are from NAOC and PMO plates. 
The magnitude differences for the common stars have a sigma value of 
about 0.20 magnitude in the B/V bands, consistent with the photometric precision of the NAOC and PMO plates.

\section{Photometric catalogs and candidates of variable sources}\label{sec:catalog}

Applying our calibration models, we obtained a total of 15,680 photometric catalogs from the calibrated plates. The catalogs include $33,282,558$ unique sources with $229,271,500$ independent observations. They are named in the format of "plate name - photometric system.fits", with the contents described in Table\,\ref{tab:column}. 

\begin{deluxetable*}{crl}[ht!] 
\tablecaption{Columns in the photometric catalogs \label{tab:column}} 
\tablehead{Number & Column name & Description}
\startdata
1-2 & ra, dec & From Gaia DR2 \\
3 & XWIN\_IMAGE & Pixel coordinates of stars on the plates \\
4 & YWIN\_IMAGE & Pixel coordinates of stars on the plates \\
5 & MAG\_AUTO & Instrumental magnitude, invalid values are set to be 99. \\
6 & OUT\_FLAG & The column ``FLAGS" in the single-exposure astrometric catalog \\
7 & XP\_FLAG & The column ``Syn\_FLAG" in the BEST database \\
8-12 & XP\_U/B/V/R/I & Standard magnitude from the BEST database \\
13-15 & DR3\_G/BP/RP & Standard magnitude from the BEST database \\
16 & excess\_factor & The column ``excess\_factor" in the BEST database \\
17 & MAG\_CORR & The magnitude after calibration \\
18 & credible\_flag & \parbox{11cm}{The values of 1, 2, 3 represent all sources involved in the zero-point calculation, the sources of $excess\_factor > 1.3+0.06 (BP-RP)^2$, and the sources whose standard magnitudes are absent, respectively.}\\
19 & variable\_flag & \parbox{11cm}{The values of 3, 4, 5, 7 and 10 represent $3\sigma, 4\sigma, 5\sigma, 7\sigma, 10\sigma$ variable candidates, respectively.}\\
20 & source\_id\_dr2 & Source id from Gaia DR2 \\
21 & source\_id\_dr3 & Source id from Gaia DR3 \\
\enddata
\end{deluxetable*}

Using the above catalogs, we have further 
selected candidates of variable stars by comparing 
the calibrated magnitudes and the standard magnitudes.
For each plate, we estimated photometric precision 
as a function of instrumental magnitude, with a sliding window of $1 {\rm \,mag}$. 
Sources that beyond the $3\sigma$ variation were regarded as 
candidates of variables. An example is shown in the 
panel (g) of Figure\,\ref{Fig:example}, where the red dots are candidates. 
We set $variable\_flag=3$ for the selected candidates. Similarly, we can also select $ 4\sigma, 5\sigma, 7\sigma, 10\sigma$ candidates, and set the $variable\_flag$ values to $4$, $5$, $7$, and $10$, 
respectively. 
The numbers of observations of $variable\_flag$ values $\geq 3, 4, 5$ are about 4.0, 1.5, and 0.66 million, respectively. While the numbers of unique variable candidates of $variable\_flag$ values $\geq 3, 4, 5$ are about 2.8, 1.1, and 0.50 million, respectively. 
Their light curves have been collected from the photometric catalog and stored separately in fits files named by Gaia DR2's source id.

\begin{figure*}[ht!] \centering
\includegraphics[width=16.cm]{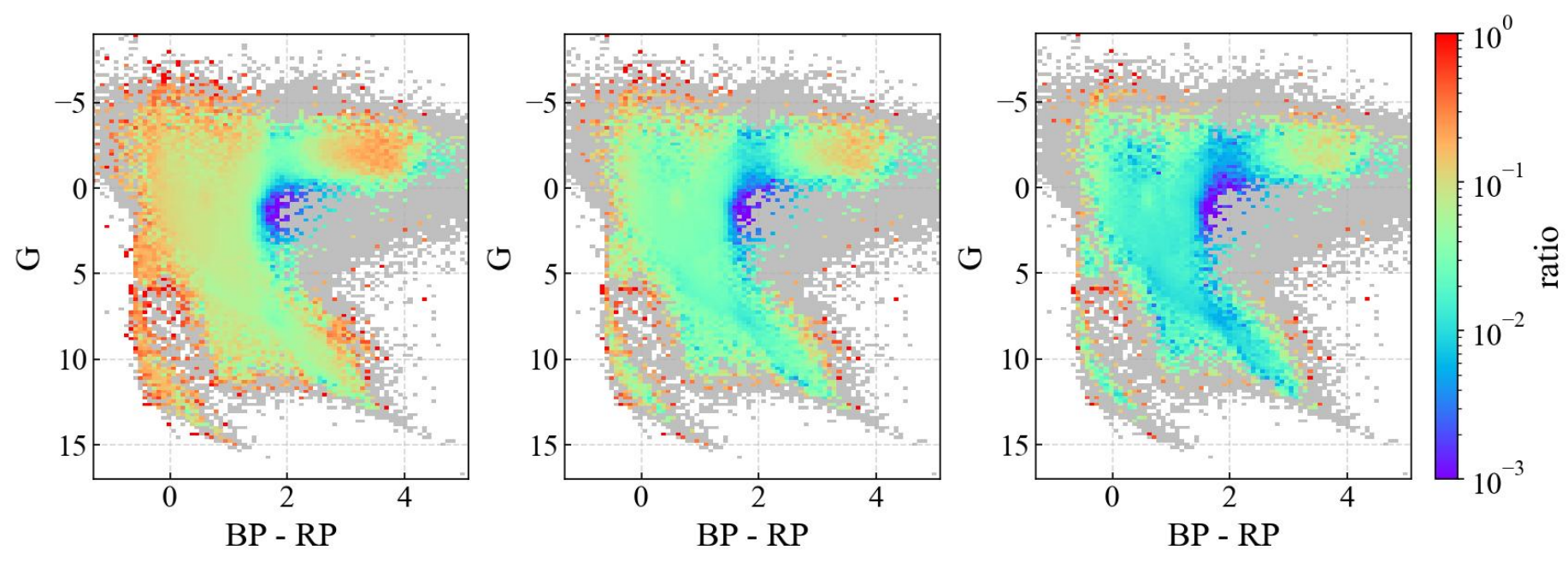} 
\caption{Fraction of variable candidates in the H-R diagram. The panels from left to right correspond to variable sources of $variable\_flag \geq 3, 4, 5$, respectively.}
\label{Fig:varhr}
\end{figure*}

The fraction of variable candidates is roughly uniform in the sky.
Additionally, we obtained the intrinsic colors and absolute magnitudes of the stars using a new three-dimensional extinction map constructed by Wang T. et al. (\url{https://doi.org/10.3847/1538-4365/adea39}) and extinction coefficients of \cite{2023ApJS..264...14Z}. 
The fraction of variable candidates in the HR diagram is shown in Figure\,\ref{Fig:varhr}. 
It is evident that the regions with higher fraction are primarily occupied by supergiants, white dwarfs, main-sequence-white-dwarfs binaries, and pre-main-sequence stars, 
as expected. 

\begin{figure*}[ht!] \centering
\includegraphics[width=17 cm]{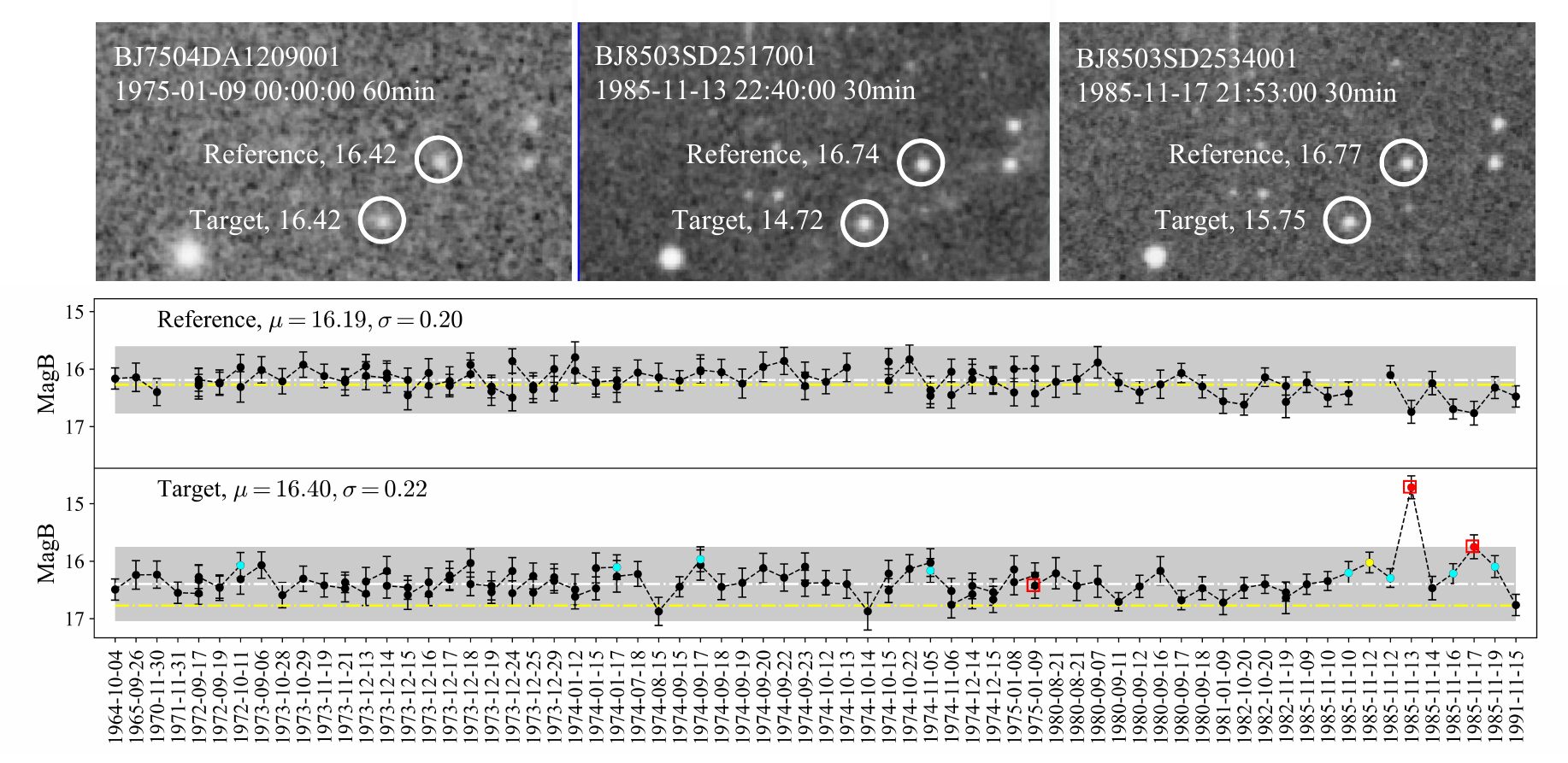} 
\caption{Images and light curves of a supergiant star ($00^{h}37^{m}20^{s}.64,~40^{\circ}16^{\prime}37^{\prime\prime}.70$)
from M\,31 and its reference star. 
The images has a field of view of approximately $3.2'\times 5.6'$, with plate name, observational time, exposure time, and calibrated magnitudes labeled. 
In the light curve panels, the blue, yellow, and red points have $variable\_flag$ values of $3$, $4$, and $5$, respectively. The $\mu$ and $\sigma$ values represent the mean and standard deviation of all calibrated magnitudes. The shaded regions cover the $3\sigma$ area. The yellow and white dashed lines represent the standard magnitudes and the mean values of calibrated magnitudes, respectively. The three red rectangles correspond to the three images above.}
\label{Fig:curve}
\end{figure*}

As an example, Figure\,\ref{Fig:curve} shows images and B-band light curves of a supergiant star in M\,31 (the target star) and its reference star.
Note that most data are from the NAOC plates. It can be seen that the significant brightening of the supergiant on November 13th and 17th, 1985, is well recorded in our data.
The standard magnitude and the mean of the calibrated magnitudes are very close for the stable reference star, as expected.   However, for the target variable star, the difference is as large as $0.4{\rm \,mag}$, which is also not surprising due to the epoch differences. 

\section{Conclusions} \label{sec:conclusion}

In this study, we employed a approach to perform photometric calibration of $15,696$ single-exposure plates
from the Chinese Plate-Digitizing Project.
Our calibration model takes into account the following effects: 1) non-linear response of the photographic plates, expressed as magnitude term; 2) difference between the real phometric system and the standard JKC system, expressed as color term; and 3) flat field to correct any position-dependent systematic errors. Using  standard stars constructed from the BEST database, we have determined the zero-point for each plate individually, modeling it independently as functions of the instrumental magnitude, Gaia $BP-RP$ color, and the source's position in the plate.
We have achieved a final photometric precision of typically 0.15, 0.23, 0.17, 0.11, 0.19 mag for SHAO, NAOC, PMO, YNAO, and QDO plates, respectively, with best cases reaching 0.07, 0.08, 0.06, 0.05, and 0.11 mag, respectively.
This method can be applied for the calibration of other digitized astronomical plates as well.

The photometric catalogs include $33,282,558$ unique sources with $229,271,500$ independent observations.
By comparing the standard magnitudes with the calibrated magnitudes, we have further identified 2.8, 1.1, and 0.50 million unique variable candidates of $variable\_flag$ values $\geq 3, 4, 5$, respectively.  Their light curves were also collected. 
Such a dataset will provide a valuable resource for conducting long-term, temporal-scale astronomical research.

All the aforementioned catalogs are publicly available on NADC (\url{https://www.doi.org/10.12149/103032}).

\vspace{2em}
We acknowledge helpful discussions with Prof. Lan Songzhu. This work is supported by the National Key R\&D Program of China (2022YFF0711500), the National Natural Science Foundation of China through the project NSFC 12222301, 12173007, 12273077, and 12403024, the Shanghai Science and Technology Innovation Action Plan (21511104100), the Global Common Challenge Special Project (018GJHZ2023110GC) and the China National Key Basic Research Program (2012FY120500).

This work has made use of data from the website of the NADC (\url{https://nadc.china-vo.org/res/r100742/}), which is supported by the Chinese Virtual Observatory (China-VO), 
and the European Space Agency (ESA) mission {\it Gaia} (\url{https://www.cosmos.esa.int/gaia}), processed by the {\it Gaia} Data Processing and Analysis Consortium (DPAC, \url{https://www.cosmos.esa.int/web/gaia/dpac/consortium}). Funding for the DPAC has been provided by national institutions, in particular the institutions participating in the {\it Gaia} Multilateral Agreement.

\clearpage
\begin{appendix}

\begin{figure*}[ht!] \centering
\includegraphics[width=12.9cm]{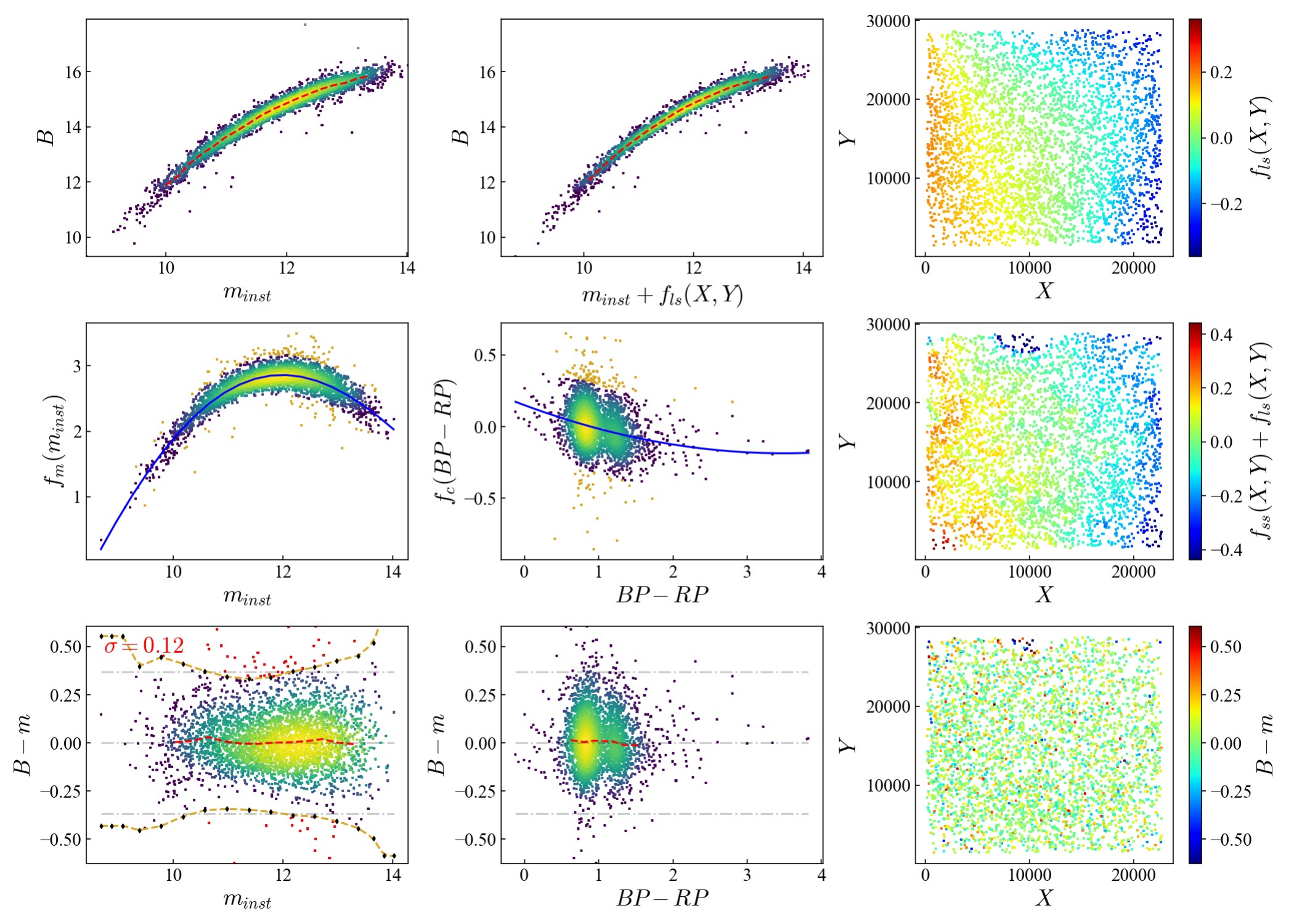} 
\caption{Same to Figure\,\ref{Fig:example} but for the plate SH6401V64087001.}
\label{Fig:sh}
\end{figure*}

\begin{figure*}[ht!] \centering
\includegraphics[width=12.9cm]{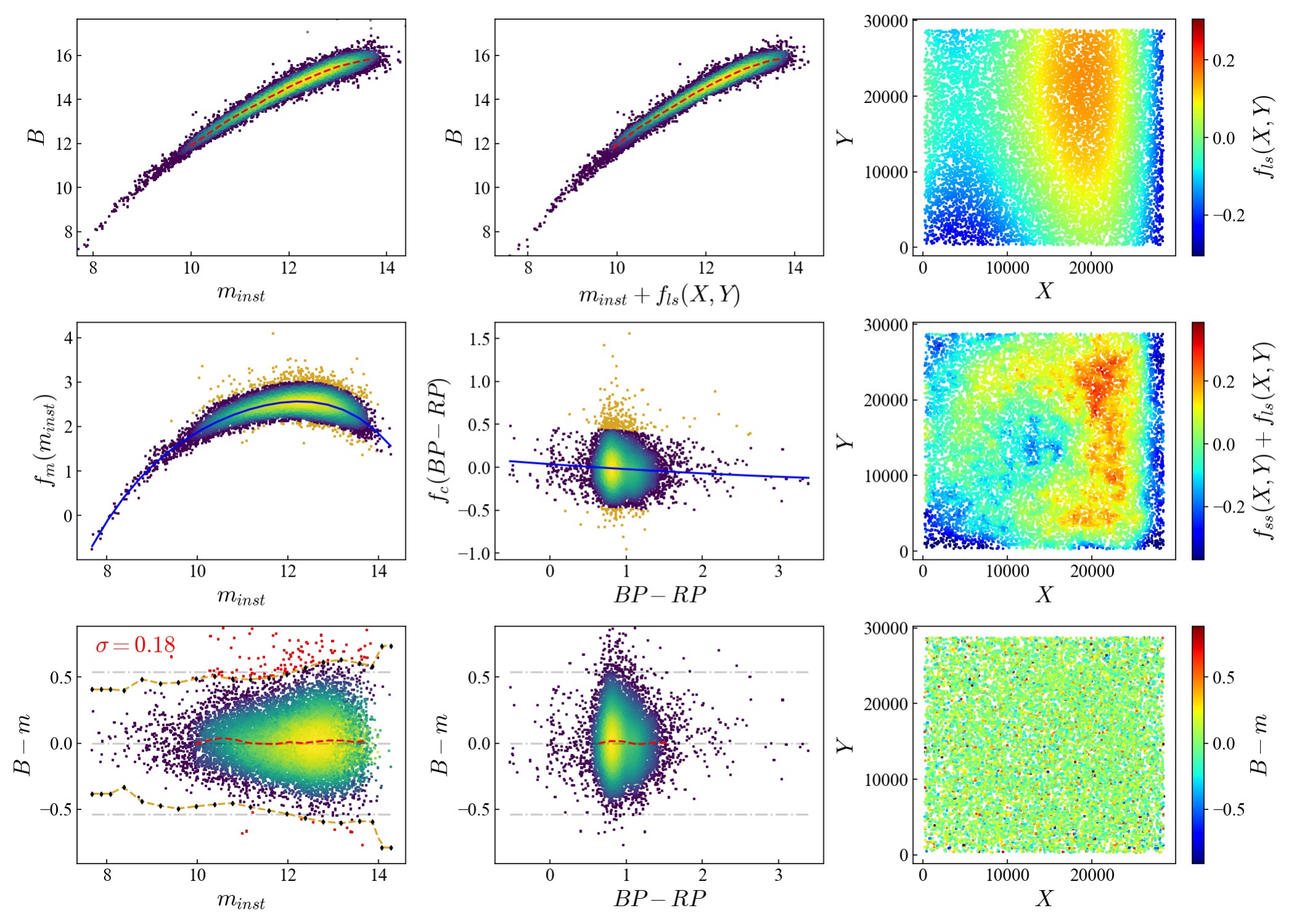} 
\caption{Same to Figure\,\ref{Fig:example} but for the plate BJ8404DA3924001.}
\label{Fig:bj}
\end{figure*}

\begin{figure*}[ht!] \centering
\includegraphics[width=12.9cm]{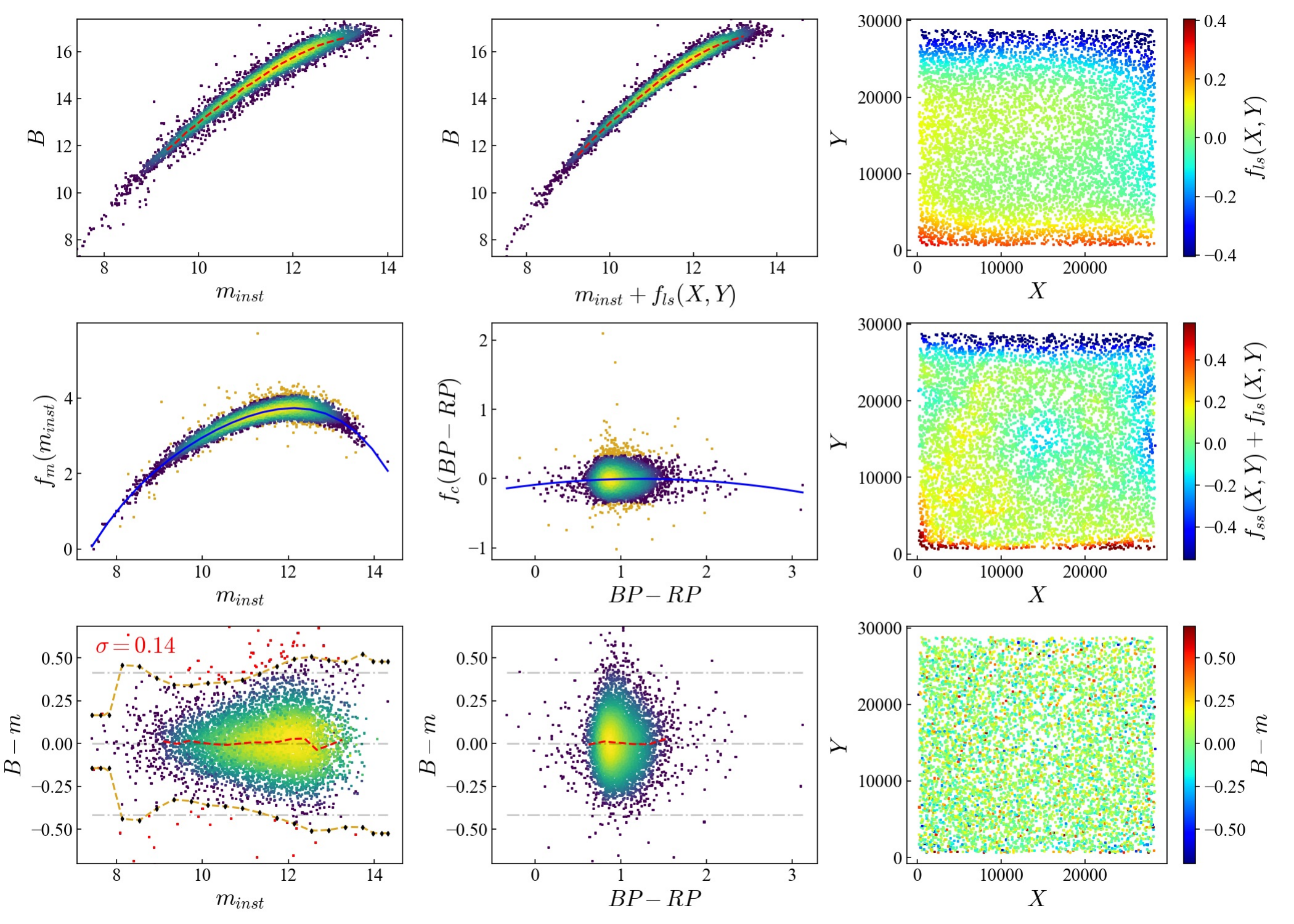} 
\caption{Same to Figure\,\ref{Fig:example} but for the plate ZT8409SN3433001.}
\label{Fig:zt}
\end{figure*}

\begin{figure*}[ht!] \centering
\includegraphics[width=12.9cm]{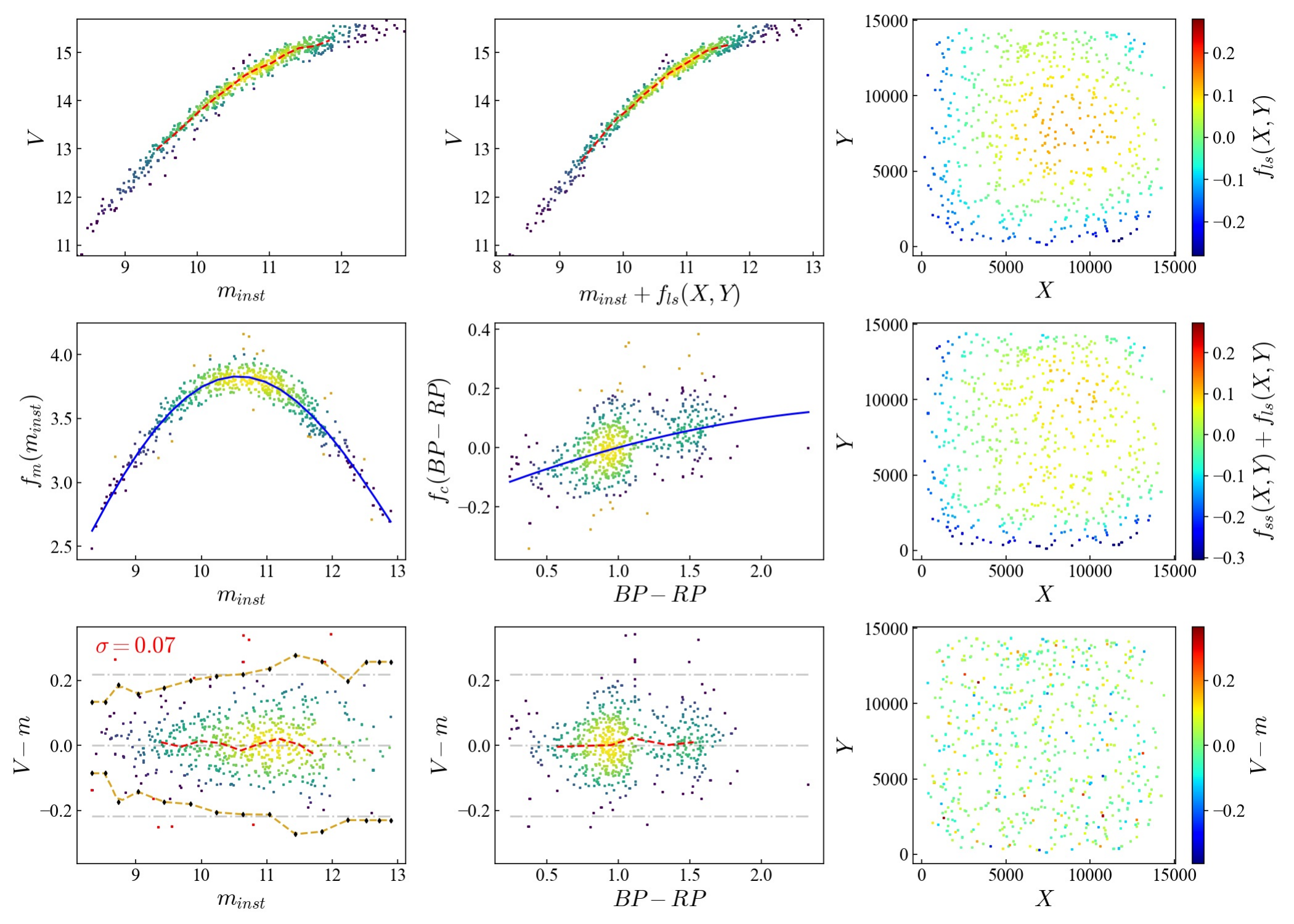} 
\caption{Same to Figure\,\ref{Fig:example} but for the plate YN8105175001.}
\label{Fig:yn}
\end{figure*}

\end{appendix}
\end{document}